\documentclass[aps,twocolumn,pra,showpacs,floatfix]{revtex4}
\usepackage{epsfig}
\usepackage{graphicx}
\usepackage{dcolumn}
\usepackage{amssymb,amsmath}
\usepackage{mathrsfs}
\begin{document}

\title{Effect of finite nuclear size on the electric quadrupole hyperfine operator}

\author{V. A. Dzuba, V. V. Flambaum}

\affiliation{School of Physics, University of New South Wales, Sydney 2052, Australia}

\begin{abstract}

We present an expression for the operator of the electric quadrupole hyperfine interaction which takes into account
finite nuclear size. We compare the results obtained with the use of this operator with those obtained in the standard approach which ignores finite nuclear size.
We found that the effect of changing operators on the hyperfine constant $B$ is small in hydrogen-like systems. 
There is a very significant  enhancement of the effect in many-electron atoms caused by the contribution of the large $s_{1/2}-d_{3/2},d_{5/2}$ and $p_{1/2}-p_{3/2}$ off diagonal matrix elements to the core polarisation, correlation and configuration interaction  corrections.
Similar enhancement takes place for transition amplitudes induced by the electric quadrupole hyperfine interaction.

\end{abstract}

\maketitle

\section{Introduction}

The study of the hyperfine structure (hfs) in heavy and superheavy atomic systems is a valuable tool for obtaining the information about nuclei~\cite{Sobelman,Johnson}.
Comparing experimental hfs with theoretical calculations of the magnetic dipole hfs constants $A$ and electric quadrupole hfs constant $B$ allows extraction of the nuclear magnetic moment $\mu$ and nuclear electric quadrupole moment $Q$. Electric quadrupole  $Q$ is strongly enhanced in deformed nuclei. 
This may serve as a guidance in the search for the nuclear stability island since the nuclei in the vicinity of the island are expected to be spherical (see, e.g.~\cite{Oganessian}).

Effect of the finite nuclear size for the magnetic hfs  constant $A$ has been extensively studied in numerous publications. It is sufficient to mention the  Bohr-Weisscopf  effect where constant  $A$ is not exactly proportional to the nuclear magnetic moment since it also depends on the distribution of magnetization inside the nucleus -  see e.g.   \cite{BW1,BW2,Orozco,BW3,BW4,BW5,BW6,BW7,Roberts} and references therein.  On the other hand, we are not aware of such study for the electric quadrupole hfs constant. 

\section{Electric quadrupole operator}

Standard operator of the electric quadrupole hfs interaction  (the $Q_{20}$ component) has the form~\cite{Sobelman,Johnson}
\begin{equation} \label{e:textbook}
\hat Q = Y_{20}/r^3,
\end{equation}
where $Y_{20}$ is the spherical function, $r$ is the distance to the nucleus. We omit here a coefficient which does not play any role in further discussion since we will discuss relative corrections to the hfs constant $B$, i.e. $\delta B/B$. 
 From the symmetry of the problem we conclude that the quadrupole electric field in the centre of the nucleus vanishes, ${\bf E}= -{\bf \nabla} \phi=0$.  The vanishing gradient  means that  the quadrupole electrostatic potential near $r=0$   is $\phi \propto r^2 Y_{20}$. This leads us to a  simple analytical form of the quadrupole operator which  takes into account the finite nuclear radius $R$:
 \begin{eqnarray}\label{e:Q}
&&\hat Q = F(r)Y_{20}, \\
&&F(r) = \left\{ \begin{array}{ll} r^2/R^5, & r \leq R \\ 1/r^3, & r > R \end{array} \right. \nonumber
\end{eqnarray}

\section{Qualitative consideration of the finite nuclear size effect}

 Let us start  from a qualitative consideration  of the dependence of $B$ on the nuclear radius $R$. Integrals in the matrix elements of the singular quadrupole operator  (\ref{e:textbook}) are dominated by small distances  $r$ from the nucleus where we can neglect energy of an electron compare to the Coulomb potential  and screening of the nuclear Coulomb potential by electrons.  Solution of the radial Dirac equation in the Coulomb field for zero energy  is expressed  in terms of the  Bessel functions $J_{\gamma_j}(x)$ and $J_{\gamma_j-1 }(x)$, where $\gamma_j=\sqrt{(j+1/2)^2 -Z^2\alpha^2}$,  $j$ is the electron angular momentum, $Z$ is the nuclear charge,   $\alpha$ is the fine structure constant, $x=(8Zr/a)^{1/2}$ is the dimensionless distance variable, $a$ is the Bohr radius- see e.g. Refs. \cite{Khriplovich,FG}.  Therefore,  the  radial dependence of the charge density of an  electron may be presented as
 \begin{equation} \label{rho}
\rho(r)=\frac{C_{nlj}}{r^2} f(x), 
\end{equation}
where the normalisation constant $C_{nlj}$ may be omitted since it cancels out in  the ratio $\delta B/B$, dimensionless function $f(x)$ is expressed as  products of Bessel functions. We can present matrix element of the operator (\ref{e:textbook})  as   
\begin{eqnarray}\label{B}
&&B \sim  \int_0^{\infty} \frac{\rho(r)}{r^3} f(x)r^2dr=C_{nlj}\left(8Z/a\right)^2 I , \\
&& I= 2 \int_0^{\infty} \frac{f(x)}{x^3}dx \sim 1 \nonumber
\end{eqnarray}
Near the nucleus Bessel functions $J_{\gamma_j}(x)$ may be expanded for  $x \ll 1$, and we have \cite{Khriplovich,FG}
\begin{equation} \label{rhosmall}
\rho(r)=C_{nlj} \left(8Z/a\right)^{2 \gamma_j} r^{2(\gamma_j -1)}. 
\end{equation}
If we use operator  (\ref{e:Q}) instead of the singular operator (\ref{e:textbook}), the contribution of the area inside the nucleus is significantly suppressed and this effect produces the change in the hfs constant $B$,   
\begin{eqnarray}
&&\delta B \sim  - C_{nlj}\int_0^{R} \frac{\rho(r)}{r^3} f(x)r^2dr , \nonumber\\
&& \frac{\delta B}{B} \sim -
\left(\frac{R Z}{a}\right )^{2(\gamma_j -1)} .
\label{deltaB}
\end{eqnarray}   
We should note that Eq. (\ref{rhosmall}) for the density $\rho(r)$ is valid outside the nucleus where the nuclear Coulomb potential is equal to $V(r)=-Z e^2/r$.  However, practically all numerical calculations of hfs have actually  taken into account finite size of the nucleus in the electron wave functions. Analytical calculations of the  electron wave function  in the finite size nucleus  potential $V(r)$ (instead of the point-like potential $V(r)=-Z e^2/r$)  have been done in Refs. \cite{Khriplovich,FG}. The main difference is that in the leading term $\rho(r) \propto r^{2j-1}$ (instead of  $r^{2(\gamma_j -1)}$; the difference in power of $r$ is $\sim Z^2 \alpha^2/(j+1/2)$). Such modification  of $\rho(r)$ inside  the nucleus produces  coefficient $\sim 1$ in the estimate of
 $\delta B/B$ and does not change any conclusions. This is easy to explain since the Coulomb wave function  for $r>R$ provides boundary condition at $r=R$  for the solution inside the nucleus, therefore, the factor $\left(\frac{R Z}{a}\right )^{2(\gamma_j -1)}$ in the estimate of $\delta B/B$ appears in any case.

 The $s_{1/2}$ and $p_{1/2}$ electronic states have zero value of $B$.  Simple estimates for the  states with total angular momentum $j>1/2$ gives $\delta B/B$  
equal to a  small fraction of per cent.  Indeed, power of the small parameter $RZ/a$  in Eq. (\ref{deltaB}) is positive, from 2 for small $Z\alpha$ to 1.46 for $Z$=137. However, this naive  estimate is only valid for the hydrogen-like single electron atoms.

In many-electron atoms the core polarization corrections and other correlation corrections contain large non-diagonal matrix elements of the hyperfine interaction such as $\langle s_{1/2}|\hat Q |d_{3/2}\rangle$,  $\langle s_{1/2}|\hat Q |d_{5/2}\rangle$  and 
$\langle p_{1/2}|\hat Q |p_{3/2}\rangle$.   These large non-diagonal matrix elements  may strongly enhance  effects of configuration mixing on $B$. They are also responsible for the transition amplitudes induced by electric quadrupole hyperfine interaction - see, for example Ref.~\cite{hfs-tr}, where probabilities of E3 and M2 atomic clock transitions, which are transformed to E1 by the hfs operators, have been calculated. Electron wave functions  $s_{1/2}$ and  $p_{1/2}$ tend to infinity for point-like nucleus and this significantly increases the sensitivity to the nuclear size: 
\begin{equation} \label{deltaBs}
\frac{\delta B}{B} \sim 
\left(\frac{R Z}{a}\right )^{\gamma_{1/2} + \gamma_{3/2}  -2}
\end{equation}
 Power of the small parameter $RZ/a$ becomes negative   for $Z>132$ (this means "infinite" $\delta B/B$ for $R=0$). However, the ratio $\delta B/B$ may exceed 1\% already for $Z>80$.
 Therefore, we should use a more accurate electric quadrupole operator  (\ref{e:Q}) inside the nucleus. Below we complement our rough  estimates by the accurate numerical calculations.
 
 \section{Hydrogen-like systems} 
 
We start our study from the hydrogen-like systems. We use the Fermi distribution of the electric charge over the nuclear volume with $R=1.2 A^{1/3}$, here $A$ is the number of nucleons in the nucleus. 
The same nuclear radius is used in (\ref{e:Q}). We perform calculations of the allowed diagonal and non-diagonal matrix elements of operator $\hat Q$ for  the $3s$, $3p_{1/2}$, $3p_{3/2}$, $3d_{3/2}$ and $3d_{5/2}$ states.
 Note that all single-electron states of the same symmetry are proportional to each other on short distances, therefore, states with any principal quantum number $n$ can be used in the study. We choose $n=3$ just for the convenience. 
The calculations are done for a set of different values of nuclear charge $Z$ and for two forms of the operator $\hat Q$, (\ref{e:textbook}) and (\ref{e:Q}).
The results are compared in Table~\ref{t:hlike}. The results are presented as a difference in per cent between the two forms of the operator. One can see that the difference is small for the diagonal matrix elements. It reaches  $\sim 10^{-3}$ for $Z=120$. 
Note also that the effect is practically zero for states with $j>3/2$. However, the effect is much larger for the off-diagonal matrix elements involving $s_{1/2}$ or $p_{1/2}$ states. This is because these states penetrate inside the nuclei. The effect reaches $\sim$~1\% for $Z=83$ (Bi atom) and becomes even larger for higher $Z$ (see Table~\ref{t:hlike}).

\maketitle

\begin{table*}
  \caption{\label{t:hlike}
    The effect of changing the electric quadrupole operator (in per cent) in matrix elements for the hydrogen-like single-electron wave functions. The numbers in last five columns are $100(m_0/m_1-1)$, where $m_0$ is the matrix element calculated with the textbook formula (\ref{e:textbook}) and $m_1$ is the matrix element calculated with the corrected formula (\ref{e:Q}), $R$ is the  nuclear radius in fm.
}
\begin{ruledtabular}
\begin{tabular}   {rrr ccccc}
\multicolumn{1}{c}{$Z$}&
\multicolumn{1}{c}{$A$}&
\multicolumn{1}{c}{$R$}&

\multicolumn{1}{c}{$s_{1/2}-d_{3/2}$}&
\multicolumn{1}{c}{$s_{1/2}-d_{5/2}$}&
\multicolumn{1}{c}{$p_{1/2}-p_{3/2}$}&
\multicolumn{1}{c}{$p_{3/2}-p_{3/2}$}&
\multicolumn{1}{c}{$d_{3/2}-d_{3/2}$}\\


\hline

  10 &   21 &  3.31071 &  0.0103 &  0.0647 &  0.0001 &  0.0000 &  0.0000  \\
  20 &   43 &  4.20408 &  0.0340 &  0.1118 &  0.0005 &  0.0001 &  0.0000  \\
  30 &   67 &  4.87386 &  0.0699 &  0.1664 &  0.0022 &  0.0004 &  0.0000  \\
  40 &   91 &  5.39753 &  0.1220 &  0.2341 &  0.0066 &  0.0009 &  0.0000  \\
  50 &  119 &  5.90242 &  0.1985 &  0.3296 &  0.0167 &  0.0020 &  0.0000  \\
  60 &  145 &  6.30431 &  0.3067 &  0.4538 &  0.0371 &  0.0037 &  0.0001  \\
  70 &  171 &  6.66060 &  0.4644 &  0.6306 &  0.0772 &  0.0067 &  0.0003  \\
  80 &  199 &  7.00593 &  0.6936 &  0.8877 &  0.1537 &  0.0118 &  0.0007  \\
  83 &  239 &  7.44699 &  0.8075 &  1.0685 &  0.1953 &  0.0150 &  0.0010  \\
  92 &  235 &  7.40521 &  1.1079 &  1.3601 &  0.3354 &  0.0223 &  0.0018  \\
 100 &  245 &  7.50879 &  1.4816 &  1.7606 &  0.5419 &  0.0326 &  0.0032  \\
 120 &  295 &  7.98832 &  3.0962 &  3.6355 &  1.7239 &  0.0879 &  0.0128  \\

\end{tabular}			
\end{ruledtabular}
\end{table*}

\begin{figure}
\epsfig{figure=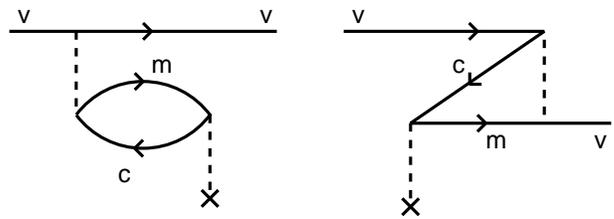,scale=0.7}
\caption{First order core polarisation correction. Cross stands for the electric quadrupole operator, $v$ is valence state, $c$ is a state in the core, $m$ is the virtual state above the core.}
\label{f:rpa}
\end{figure}

\begin{table*}
 \caption{\label{t:atoms}
    The effect of changing the electric quadrupole operator (in per cent) in the diagonal matrix elements for the valence single-electron wave functions of many-electron atoms and ions. The numbers in the last five columns are $100(m_0/m_1-1)$, where $m_0 = \langle v|\hat Q |v \rangle$ is the matrix element calculated with the textbook formula (\ref{e:textbook}) and $m_1$ is the matrix element calculated with the correct formula (\ref{e:Q}). $n=7$ for Fr, U, Fm and No, $n=8$ for E120.
}
\begin{ruledtabular}
\begin{tabular}   {rl ccccc}
\multicolumn{1}{c}{$Z$}&
\multicolumn{1}{c}{Atom}&

\multicolumn{1}{c}{$v=np_{3/2}$}&
\multicolumn{1}{c}{$v=(n-1)d_{3/2}$}&
\multicolumn{1}{c}{$v=(n-1)d_{5/2}$}&
\multicolumn{1}{c}{$v=(n-2)f_{5/2}$}&
\multicolumn{1}{c}{$v=(n-2)f_{7/2}$}\\

\hline
 87 & Fr~I    & 0.059 & 0.037 & 0.216 & 0.162 & 0.165 \\
 92 & U~VI    & 0.041 &  -0.020 & 0.070 & 0.073 & 0.088 \\
100 & Fm~I    &  &  &  &  0.069 & 0.072 \\
102 & No~II   & 0.108 & -0.090 &  0.392 &  &  \\
120 & E120~II & 0.433 &  -1.373 & 1.664 & 1.315 & 1.526 \\

\end{tabular}			
\end{ruledtabular}
\end{table*}

\section{Many-electron atoms} 

Next, we study the effect of changing the electric quadrupole operator $\hat Q$ from (\ref{e:textbook}) to (\ref{e:Q}) on the diagonal matrix elements  (i.e. on the electric quadrupole constants $B$) in many-electron atoms.
As examples, we consider heavy atoms or ions with a relatively simple electronic structure, one electron above closed shells. In atoms with several valence electrons the effect may be even bigger due to the large configuration mixing, which involves  the  $s_{1/2}-d_{3/2},d_{5/2}$ and $p_{1/2}-p_{3/2}$  matrix elements of  $\hat Q$ .
The calculations are done in the $V^{N-1}$ approximation, which means that the initial relativistic Hartree-Fock (RHF) calculations are performed for the closed-shell core, the states of external electron are calculated in the field of the frozen core. To calculate matrix elements of the $\hat Q$ operator, we use the time-dependent Hartree-Fock method which is equivalent to the well-known random-phase approximation (RPA) - see  e.g.  ~\cite{Johnson}. 
 The RPA equations can be written as (see  e.g ~\cite{TDHF}) 
\begin{equation}
(\hat H_0 - \epsilon_c)\delta \psi_c = - (\hat Q + \delta V)\psi_c.
\label{e:RPA}
\end{equation}
Here $\hat H_0$ is the RHF operator, index $c$ numerates states in the core, $\psi_c$ is a single-electron wave function for a particular state in the core, $\delta \psi_c$ is the correction to $\psi_c$ caused by external field $\hat Q$, $\delta V$ is the correction to the self-consisted RHF potential caused by the change in all core wave functions. The RPA equations (\ref{e:RPA}) are solved self-consistently for all states in the core to find $\delta V$.
Matrix elements for valence states $v$ are then calculated as  \\$\langle v | \hat Q + \delta V | v \rangle$.

The results of calculations are presented in Table~\ref{t:atoms}. We included Fr and U as heavy atoms of a broad experimental interest. We also included Fm and No as heaviest atoms for which atomic spectra  measurements are in progress~\cite{Fm1,Fm2,No1,No2,No3,No4}. We included E120 for illustration on how big the effect could be for very high $Z$. We see from the table that the effect  on $B$ in many-electron atoms is significantly larger than that  for the hydrogen-like systems (see Table~\ref{t:hlike}). This is due to the contribution of the $s_{1/2}-d_{3/2},d_{5/2}$ and $p_{1/2}-p_{3/2}$ off-diagonal matrix elements into the core polarisation correction (see Fig.~\ref{f:rpa}).

\begin{table}
 \caption{\label{t:CP}
Relative values of the core polarisation correction to the matrix elements of the valence states of E120~II as well as their decomposition over different core channels (all numbers in per cent). Core polarisation correction $\langle v| \delta V| v \rangle$ is related to the total matrix element $\langle v| \hat Q +\delta V| v \rangle$.
One channel is the sum over all core states of given type ($s_{1/2}$, $p_{1/2}$, etc.) and all possible excited states in the expression for core polarisation (see diagrams on Fig.~\ref{f:rpa}).
}
\begin{ruledtabular}
\begin{tabular}   {l rrrrr}
&\multicolumn{1}{c}{$v=8p_{3/2}$}&
\multicolumn{1}{c}{$v=7d_{3/2}$}&
\multicolumn{1}{c}{$v=7d_{5/2}$}&
\multicolumn{1}{c}{$v=6f_{5/2}$}&
\multicolumn{1}{c}{$v=6f_{7/2}$}\\
\hline
 &\multicolumn{5}{c}{Relative CP correction (per cent)}\\
    &       36 &   20  &   60 & 91 & 93 \\  
    \hline  
\multicolumn{1}{c}{Channel}&
\multicolumn{5}{c}{Decomposition over channels (per cent)}\\

$s_{1/2}$ &   20 &   304 &   82 &   59 &   52 \\
$p_{1/2}$ & -158 & -1505 & -579 & -349 & -361 \\
$p_{3/2}$ &  255 &  1487 &  652 &  439 &  450 \\
$d_{3/2}$ &  -32 &  -373 & -117 &  -79 &  -73 \\
$d_{5/2}$ &   15 &   186 &   63 &   29 &   31 \\
$f_{5/2}$ &   -3 &   -31 &  -10 &   -5 &   -5 \\
$f_{7/2}$ &    3 &    32 &    9 &    6 &    6 \\
Total     &  100 &   100 &  100 &  100 &  100 \\

\end{tabular}			
\end{ruledtabular}
\end{table}

For better understanding of the role of the off-diagonal matrix elements in the core polarisation correction we present in Table~\ref{t:CP} the decomposition of the corrections to matrix element of different states of E120$^+$ over different channels in the core. One channel is the sum over all core states $c$ of particular symmetry and all possible states $m$ above the core. For example, $s$-channel  contains terms with the $\langle 1s|\hat Q+\delta V|nd_{3/2}\rangle$,  $\langle 1s|\hat Q+\delta V|nd_{5/2}\rangle$, $\langle 2s|\hat Q+\delta V|nd_{3/2}\rangle$, etc. The $s_{1/2}$ and $p_{1/2}$ channels give non-zero contribution due to off-diagonal matrix elements only. The off-diagonal matrix elements contribute to other channels too. For example, the $\langle 2p_{3/2}|\hat Q+\delta V|np_{1/2}\rangle$ matrix elements contribute to the $p_{3/2}$ channel.  

As can be seen from Table~\ref{t:CP}, the contribution of the off-diagonal matrix elements is huge (mostly, in channels $s_{1/2}$ and $p_{1/2}$)  Some contributions of the off-diagonal matrix elements exceed many times the final answer since there are partial cancellations of these big contributions. Large off-diagonal matrix elements are more sensitive to the form of the operator $\hat Q$, see Table~\ref{t:hlike}. 
These two facts lead to significant enhancement of the effect in many-electron atoms.

Calculations of the probabilities of clock transitions induced by the hyperfine interaction (see, for example Ref.~\cite{hfs-tr}) can also benefit from considering correct form of the electric quadrupole operator $\hat Q$. These transitions, forbidden as electric dipole transitions, are open by off-diagonal mixing of states with different  electron angular momentum  by the magnetic dipole or electric quadrupole interaction~\cite{hfs-tr}. 

\vspace{3mm}
\textit{Acknowledgements} --- 
The work was supported by the Australian Research Council Grants No.\ DP230101058 and DP200100150.

\end{document}